\documentclass[12pt]{article}
\usepackage{amsfonts,amssymb,amsmath,epsfig,graphicx}
\usepackage{theorem,calc}
\usepackage{hhline}

\newtheorem{definition}{Definition}[section]

\newtheorem{proposition}[definition]{Proposition}

\numberwithin{equation}{section}

\newcommand{\beq}{\begin{equation}}
\newcommand{\eeq}{\end{equation}}
\newcommand{\bea}{\begin{eqnarray}}
\newcommand{\eea}{\end{eqnarray}}
\newcommand{\beano}{\begin{eqnarray*}}
\newcommand{\eeano}{\end{eqnarray*}}

      \def\cB{{\cal B}}

            \def\cR{{\cal R}}

\newcommand{\CC}{{\mathbb C}}

\newcommand{\PP}{\mbox{${\mathbb P}$}}

\newcommand{\yR}{\mbox{$ \cal R  $}}
\newcommand{\rR}{\mbox{$ \cal B  $}}

\newcommand{\prf}{\underline{Proof:}\ }
\newcommand{\finprf}{\null \hfill {\rule{5pt}{5pt}}\\[2.1ex]\indent}
\newcommand{\ie}{{\it i.e.}\ }

\begin{document}

\setcounter{page}{0}
\pagestyle{empty}

\null
\begin{center}

{\LARGE \bf Set-theoretical reflection equation:\vspace{1mm}\\
Classification of reflection maps}

\vspace{1cm}

{\large V. Caudrelier$^a$, N. Cramp\'e$^{b}$ and Q.C. Zhang$^a$}\\

\vspace{0.5cm}
\emph{$^a$ Centre for Mathematical Science, City University London,\\
Northampton Square, London EC1V 0HB, UK}\\
\vspace{0.5cm}
\emph{$^b$ CNRS, Universit\'e Montpellier II, Laboratoire Charles Coulomb, UMR 5221, \\
Place Eug\`ene Bataillon - CC070, F-34095 Montpellier, France}\\
\vspace{0.5cm}

\textit{Email: v.caudrelier@city.ac.uk, nicolas.crampe@um2.fr, cheng.zhang.5@city.ac.uk}
\end{center}

\vfill

\begin{abstract}
The set-theoretical reflection equation and its solutions, the reflection maps, recently introduced by two of the authors, is presented in general and 
then applied in the context of quadrirational Yang-Baxter maps. 
We provide a method for constructing reflection maps and we obtain a classification of solutions associated to all the families of quadrirational Yang-Baxter maps 
that have been classified recently. 
\end{abstract}

\vspace{1cm}


\vfill

\newpage
\pagestyle{plain}

\section*{Introduction}
The Yang-Baxter equation, introduced to study a gas of spin with a $\delta$ interaction \cite{yang}
or a 2-dimensional solvable statistical model in \cite{baxter}, has been developed in many 
directions and found applications in different areas of mathematics and physics. Much of the effort has been 
concentrated on the so-called \textit{quantum} Yang-Baxter equation which is at the heart of the field of quantum integrable systems.
Although less well-known, the set-theoretical solutions of the Yang-Baxter equation have also attracted some attention.
This question goes back to Drinfeld \cite{Drin} (and even to \cite{Skly88} where a solution was given).
Then, different aspects have been developed and numerous connections have been established with, for example,
the geometric crystals \cite{BK,Eti},
the soliton cellular automatons \cite{TS,HKT},
the scattering of multicomponent solitons \cite{Gon,Tsu,APT},
the discrete integrable systems \cite{ABS,PTV} 
and the Lax or transfer matrices \cite{Ves,SV}. In view of these applications, there is also a lot of 
effort put into the possibility of classifying the solutions of the set-theoretical 
Yang-Baxter equations \cite{ABS2,PSTV}.

In the context of quantum integrable systems with boundaries, there exists, in addition to the Yang-Baxter 
equation, a second equation: the reflection equation.
This equation was introduced to encode the reflection on the boundary of particles 
in quantum field theory \cite{Che} and to prove the integrability of 
quantum models with boundaries \cite{Sk}.  
This equation appears also as the defining relations for subalgebras of quantum groups \cite{KS,MNO} 
which are the
symmetry of quantum models such as the $BC_n$ Sutherland models \cite{BPS,CC}, as a relation to construct 
commuting familly of time evolutions for box-ball system with reflectiong end \cite{KOY}
and as a consistency relation for 
the coordinate Bethe ansatz applied to the $\delta$ gas with boundaries \cite{HL} and impurity \cite{bart}.

Motivated by all these fields of application of the set-theoretical Yang-Baxter equation and of the reflection
equation, we propose in this article to explore the interplay between these two notions and to study 
the set-theoretical reflection equation.
Surprisingly, this equation was introduced only very recently by two of the authors \cite{VZ} 
in the context of factorization of vector soliton interactions on the half-line. There, it appeared naturally 
as the equation ensuring that the interactions of the polarizations of an $N$-soliton solution among themselves and with the 
boundary factorises consistently as a sequence of soliton-soliton and soliton-boundary interactions. At the origin of 
the study in \cite{VZ} lie the results of \cite{VZ2}. The latter falls into the vast area of integrable PDEs on the 
half-line for which various related approaches can be found in \cite{Skly_classical} (classical $r$ matrix approach), \cite{H} 
(B\"acklund transformations approach), \cite{GGH} (symmetry algebra approach) or \cite{Fokas} (inverse scattering method approach).

In the first section of this paper, after recalling the important facts about the set-theoretical Yang-Baxter (YB)
equation, we give the definition of the set-theoretical reflection equation and its solutions, the reflection maps.
We also give its parametric version and comment on its graphical representation. We discuss, in this context, the construction
of the transfer map to compare with \cite{Ves} in the case of the YB equation.
The second section is dedicated to the construction of reflection maps. The argument is based on a folding technique. 
Using this for the set-theoretical reflection equation associated to the YB maps classified in 
\cite{ABS2,PSTV}, we obtain a classification of solutions. Finally, we conclude with the numerous 
open problems and perspectives related to our study.


\section{Set-theoretical reflection equation}

We present some elements and facts of the theory of reflection maps and set-theoretical reflection equation first introduced 
at the end of the paper \cite{VZ}.
To do so, we first need to recall some definitions related to YB maps (see e.g. \cite{Ves}).

\subsection{Yang-Baxter maps}

Let $S$ be a set and ${\cal R} : S \times S \to S \times S$ a map from the Cartesian product of $S$ onto itself,
\bea
\label{eq:RR}
 {\cal R}(X, Y) = (U,V)\equiv (f (X, Y), g(X, Y))\,.
\eea
Define ${\cal R}_{ij}  : S^N \to S^N$ as the map acting as ${\cal R}$ on the $i$th and $j $th 
copies of $S$ in the $N$-fold Cartesian product $S^N$ 
and identically on the others. More precisely, for $i<j$,
\bea
{\cal R}_{ij} (X_1, \dots  ,X_n ) &=& (X_1 , \dots  ,X_{i- 1} , f (X_i , X_j ), \dots ,  g(X_i , X_j ), X_{j +1} , \dots , X_n )\,, \\ 
{\cal R}_{ji} (X_1, \dots  ,X_n ) &=& (X_1 , \dots  ,X_{i-1} , g (X_j , X_i ) , \dots ,       f(X_j , X_i ), X_{j+1} , \dots , X_n ) \,.
\eea
 In particular, for $N = 2$, ${\cal R}_{12} \equiv {\cal R}$.
If ${\cal R}$ satisfies the following Yang-Baxter relation 
\begin{equation}
  \label{eq:yba112}
    {\cal R}_{12}   {\cal R}_{13}   {\cal R}_{23} = {\cal R}_{23}    {\cal R}_{13}    {\cal R}_{12} \,, 
\end{equation}
as an identity on $S\times S\times S$, then ${\cal R}$ is called a Yang-Baxter map. In addition, if ${\cal R}$ satisfies 
\begin{equation}
  \label{eq:yba113}
     {\cal R}_{21}   {\cal R}_{12} = Id\,,
\end{equation}
then ${\cal R}$ is called a reversible Yang-Baxter map. 
       
It is useful to introduce the so-called parametric Yang-Baxter map ${\cal R}(a,b)$ which is an important special case 
obtained by considering $S\times \Sigma$, where $\Sigma$ is some parameter set (usually $\CC$), instead of only $S$ as before. 
The corresponding map
\bea
\label{eq:param}
\cR:(X,a;Y,b)\mapsto (f(X,a;Y,b),a;g(X,a;Y,b),b)\equiv(f_{ab}(X,Y),a;g_{ab}(X,Y),b)\,,
\eea
is conveniently written as $\cR(a,b)$ and seen as acting only on $S\times S$:
\bea
\cR(a,b):(X,Y)\mapsto (f_{ab}(X,Y),g_{ab}(X,Y))\,.
\eea
It satisfies the parametric YB equation
\begin{equation}
  \label{eq:yba114}
   {\cal R}_{12}(a,b)   {\cal R}_{13}(a,c)   {\cal R}_{23}(b,c) = {\cal R}_{23} (b,c)   {\cal R}_{13} (a,c)  {\cal R}_{12}(a,b)\,,
\end{equation}
and the corresponding reversibility condition reads
\begin{equation}
  \label{eq:yba115}
   {\cal R}_{21}(b,a)   {\cal R}_{12} (a,b) = Id\,.
\end{equation}
The notion of parametric maps is important, for example, for YB maps arising from the study of vector or matrix soliton solutions of certain classical integrable 
nonlinear equations \cite{Tsu,APT,Gon}.

\begin{figure}[htb]
\begin{center}
\includegraphics[width=.3\textwidth]{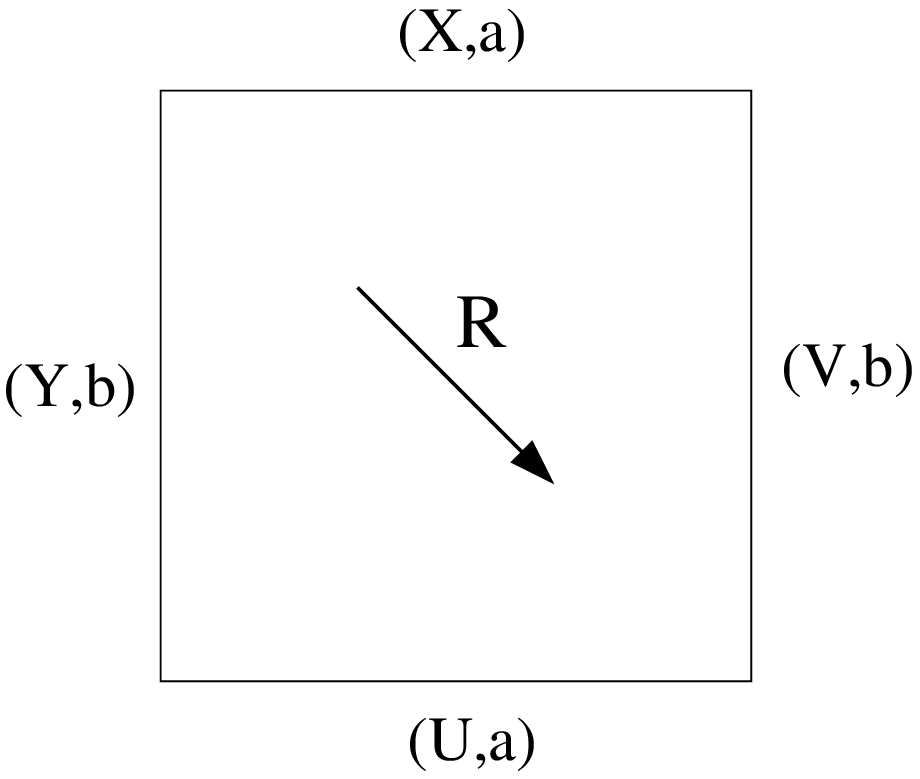}
\caption{Map $\yR$}\label{fig:Rmap}
\end{center}
\end{figure}

\begin{figure}[htb]
\begin{center}
\includegraphics[width=.7\textwidth]{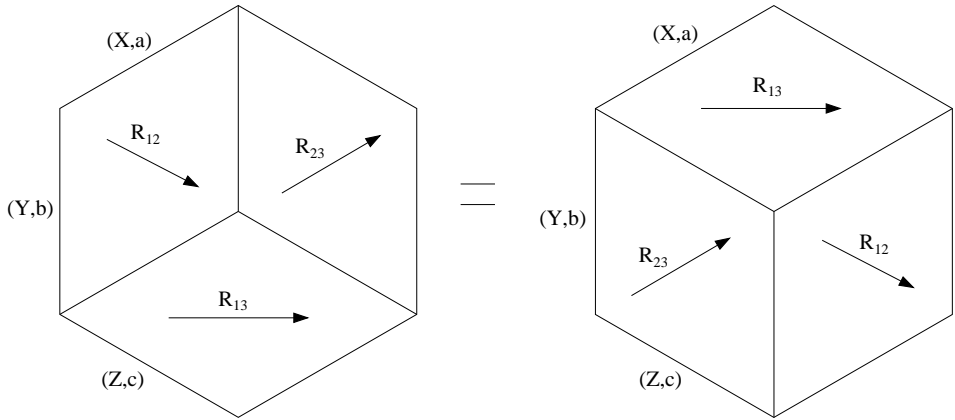}
\caption{Graphical representation of the set-theoretical Yang-Baxter equation}\label{fig:ybe}
\end{center}
\end{figure}

There is a well-known pictorial description of YB maps (see e.g. \cite{ABS2}). A map $\yR$ defined by (\ref{eq:param}) is associated 
to a quadrilateral as in Fig. \ref{fig:Rmap}.
Then, the parametric YB equation (\ref{eq:yba114}) can be understood as the consistency condition ensuring that the two 
possible orders of action of the map $\yR$ on $(X,a;Y,b;Z,c)$, described by both sides of the equation shown in Fig.\ref{fig:ybe}, provide the same result.
Let us also remark the similarities between this graphical representation of the set-theoretical YB equation and the 
so-called face representation of the YB equation \cite{AK,FHS,BPO}.

\subsection{Reflection maps}

Based on the form of the quantum reflection equation \cite{Che,Sk} and on the important role played by its set-theoretical version in \cite{VZ}, 
we now introduce the following general definition of a reflection map.
\begin{definition} Given a Yang-Baxter map $\yR$, a reflection map $\rR$ is a solution of the set-theoretical reflection equation
\begin{eqnarray}
\rR_1 \yR_{21} \rR_2 {\cal R}_{12} =\yR_{21} \rR_2 \yR_{12}   \rR_1\,,
\end{eqnarray}
as an identity on $S\times S$.
The reflection map is called involutive if 
\begin{eqnarray}
\rR\rR=Id\,.
\end{eqnarray}
\end{definition}
To introduce the notion of parametric reflection map, we need some extension of the above discussion for parametric YB maps in order 
to encompass known situations. Indeed, a parametric YB map has a trivial action on the parameter set $\Sigma$ but 
it is known from the solutions found in \cite{VZ} for reflection maps that a parametric reflection map can in general have a 
nontrivial action on the parameter set. In general this reads
\begin{equation}
\label{def_ref_map}
 {\rR} : (X, a ) \mapsto   (h_a(X), \sigma(a))~,~~X\in S~,~~a\in\Sigma\,,
\end{equation}
for some maps $h_a$ and $\sigma$.
One can then use the convenient notation $\rR(a)$ for a parametric reflection map but one has to keep in mind the nontrivial action of
$\sigma$ when composing a reflection map with other maps. To illustrate this, we give the definition of a parametric reflection map as follows:
\begin{definition}
Given a parametric Yang-Baxter map $\yR_{12}(a,b)$, $a,b\in \Sigma$ and a map $\sigma:\Sigma\to \Sigma$, 
a parametric reflection map $\rR(a)$ is a solution of the parametric set-theoretical reflection equation
\begin{equation}
\label{param_ref_eq}
\rR_1(a) \yR_{21}(\sigma(b),a) \rR_2(b) {\cal R}_{12}(a,b) =\yR_{21}(\sigma(b),\sigma(a)) \rR_2(b) \yR_{12}(\sigma(a),b)   
\rR_1(a)
\end{equation}
as an identity on $S\times S$. The reflection map is called involutive if $\sigma$ is an involution in $\Sigma$ and
\begin{eqnarray}
\rR(\sigma(a))\rR(a)=Id\,.
\end{eqnarray}
\end{definition}
Note that classes of parametric reflection maps were constructed in \cite{VZ} for $S=\CC\PP^{n-1}$, $\Sigma=\CC$ and 
$\sigma(k)=-k^*$ with
${\cal R}_{12}(k_1,k_2)$ being the parametric YB map corresponding to the vector nonlinear Schr\"odinger equation.

\begin{figure}[htp]
\begin{center}
\includegraphics[width=.2\textwidth]{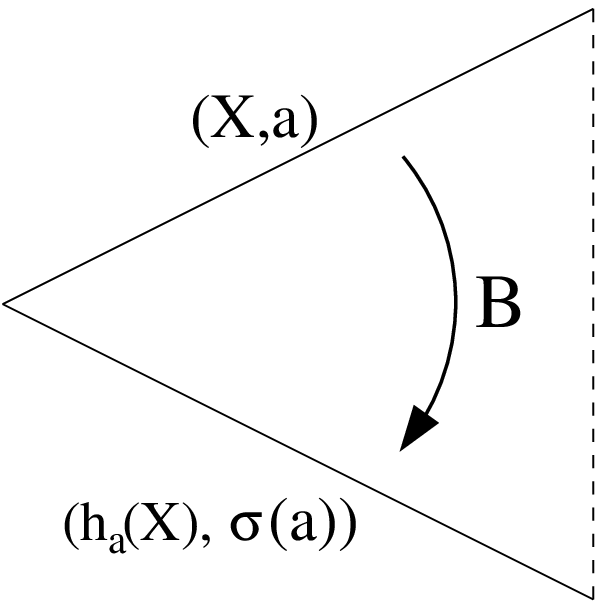}
\caption{Map $\rR$}\label{fig:Kmap}
\end{center}
\end{figure}

\begin{figure}[htp]
\begin{center}
\includegraphics[width=.7\textwidth]{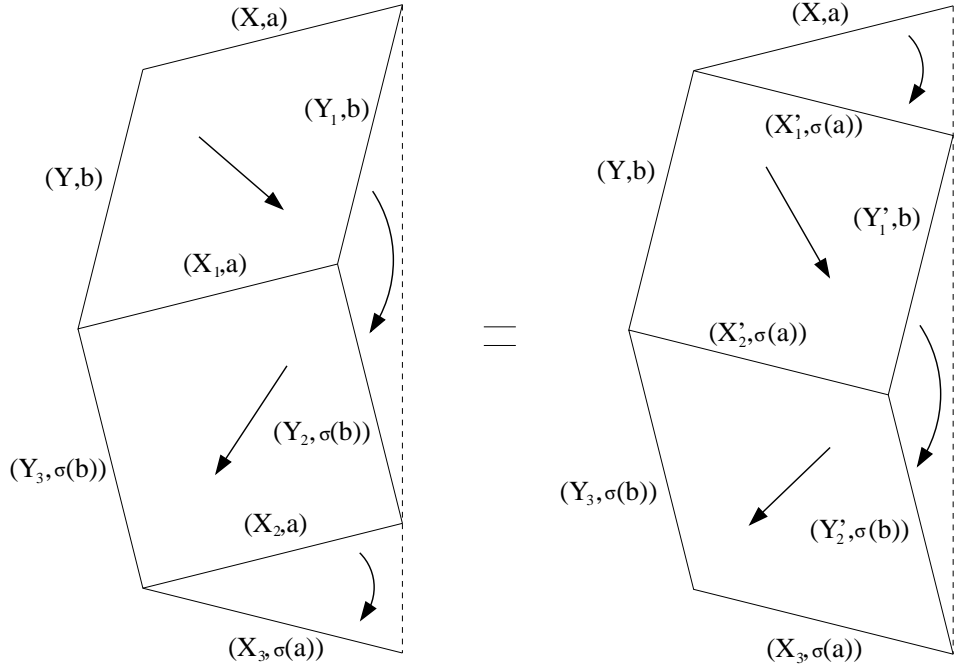}
\caption{Graphical representation of the reflection equation}\label{fig:REmap}
\end{center}
\end{figure}

Analogous to the quadrilateral description of a YB map, we introduce a pictorial description of a reflection map as a 
half-quadrilateral shown in Fig. \ref{fig:Kmap}.
The set-theoretical reflection equation can then be depicted as in Fig. \ref{fig:REmap}.
Let us remark that this visualization is reminiscent of the face representation of the reflection equation in \cite{AK,FHS,BPO}. There are also
similitudes between Fig. \ref{fig:REmap} and half of the figure representing the tetrahedron equation for 
interaction round-a-cube \cite{Bax86,BMS}.
Explicitly, the left-hand side represents the following chain of maps
\begin{eqnarray}
 (X,a;Y,b) \xrightarrow{\yR_{12}}(X_1,a;Y_1,b)  
\xrightarrow{\rR_{2}} (X_1,a;Y_2,\sigma(b)  ) 
\xrightarrow{\yR_{21}} (X_2,a;Y_3,\sigma(b)) 
\xrightarrow{\rR_{1}} (X_3,\sigma(a);Y_3,\sigma(b)) \nonumber
\end{eqnarray}
with
\begin{eqnarray}
 &&X_1=f_{ab}(X,Y)~,~~Y_1=g_{ab}(X,Y)~,~~Y_2=h_b(Y_1) \,,\nonumber\\
 &&X_2=g_{\sigma(b)a}(Y_2,X_1)~,~~Y_3=f_{\sigma(b)a}(Y_2,X_1)~,~~X_3=h_a(X_2)\,.\nonumber
\end{eqnarray}
Similarly, the right-hand side means
\begin{eqnarray}
 (X,a;Y,b) \xrightarrow{\rR_{1}} (X_1',\sigma(a);Y,b)  
\xrightarrow{\yR_{12}} (X_2',\sigma(a);Y_1',b) 
\xrightarrow{\rR_{2}}  (X_2',\sigma(a);Y_2',\sigma(b)) 
\xrightarrow{\yR_{21}} (X_3,\sigma(a);Y_3,\sigma(b)) \nonumber
\end{eqnarray}
with 
\begin{eqnarray}
&&X_1'=h_a(X)~,~~ X_2'=f_{\sigma(a)b}(X_1',Y)~,~~Y_1'=g_{\sigma(a)b}(X_1',Y)\,, \nonumber\\ 
&& Y_2'=h_b(Y_1')~,~~ X_3=g_{\sigma(b)\sigma(a)}(Y_2',X_2' )~,~~Y_3=f_{\sigma(b)\sigma(a)}(Y_2',X_2' )\,.\nonumber
\end{eqnarray}
So, Eq. (\ref{param_ref_eq}) ensures that the two ways of obtaining $(X_3,Y_3)$ from $(X,Y)$ lead to the same result.

\subsection{Transfer maps}

We conclude this presentation on the theory of reflection maps by defining the notion of transfer maps in analogy with that 
introduced in \cite{Ves}. 
Fix $N\ge 2$ and define for 
$j=1,\dots, N$ the following maps of $S^N$ into itself,
\begin{eqnarray}
\label{transfer_map}
{\cal T}_j={\cal R}_{j+1j}\dots {\cal R}_{Nj}\rR_j^-{\cal R}_{jN}\dots{\cal R}_{jj+1}
{\cal R}_{jj-1}\dots{\cal R}_{j1}\rR_j^+{\cal R}_{1j}\dots {\cal R}_{j-1j}\,,
\end{eqnarray}
where $\rR^+$ is a solution of 
\begin{eqnarray}
\label{RE_B+}
\rR_1 \yR_{21}\rR_2 {\cal R}_{12}=\yR_{21} \rR_2 \yR_{12} \rR_1\,,
\end{eqnarray}
and $\rR^-$ a solution of 
\begin{eqnarray}
\rR_1 \yR_{12}\rR_2 {\cal R}_{21}=\yR_{12} \rR_2 \yR_{21} \rR_1\,.
\end{eqnarray}
Then one proves by direct (but long) calculation the following result
\begin{proposition}
For any reversible Yang-Baxter map $\yR$, the transfer maps \eqref{transfer_map} commute with each other
\begin{eqnarray}
{\cal T}_j{\cal T}_\ell={\cal T}_\ell{\cal T}_j~,~~j,\ell=1,\dots,N\,.
\end{eqnarray}
\end{proposition}
Of course, this result transfers immediately to the case of parametric maps.

\section{Reflection maps for quadrirational Yang-Baxter maps}

In this section, we provide a classification of solutions for reflection maps associated to the $10$ families of quadrirational YB maps 
found in \cite{ABS2,PSTV}. To achieve this, we consider a certain class of functions of rational type and use an adaptation of the 
folding argument used in \cite{VZ}. This is further combined with a duality property which we derive from the particular properties
of quadrirational maps exhibited in \cite{ABS2}. The overall amount of work required is reduced by applying the notion of symmetry as explained below.

\subsection{Symmetry and quadrirational Yang-Baxter maps}

We first need the following proposition shown in \cite{PSTV}.

\begin{proposition} 
Let $s(a)$ be an involutive symmetry of the parametric Yang-Baxter map $ {\cal R}(a,b)$ i.e.
\begin{equation}
 s(a)^2=Id \quad\text{and}\quad s(a)\times s(b)\ {\cal R}(a,b)={\cal R}(a,b)\ s(a)\times s(b)\;.
\end{equation}
The map defined by
\begin{equation}\label{eq:Rs}
{\cal R}^s(a,b)=\big(s(a)\times Id\big)\ {\cal R}(a,b)\ \big(Id\times s(b)\big)
\end{equation}
is a Yang-Baxter map.
\end{proposition}
Such symmetry maps $s(a)$ allow one to relate YB maps that look different a priori. This is in particular the case for quadrirational YB maps 
\ie maps of a certain rational form acting on $S=\CC\PP^1$ with parameter set $\Sigma=\CC$: the $F$ families and 
the $H$ families (in the notations of 
\cite{ABS2,PSTV}) are related by such symmetry maps. In general, in the context of YB maps, different maps
may indeed be related to actual different underlying systems. But in the context of reflection maps, the following proposition shows that 
symmetries leave the set of solutions invariant.
\begin{proposition}
The map $\cB(a)$ is a parametric reflection map for the parametric Yang-Baxter map ${\cal R}(a,b)$ if and only if
the map $\cB(a)$ is a parametric reflection map for the parametric Yang-Baxter map ${\cal R}^s(a,b)$.
\end{proposition}
\prf
The proof of this proposition consists in writing the set-theoretical reflection equation for ${\cal R}^s(a,b)$ expressed thanks to (\ref{eq:Rs}).
Then, remarking for example that $\big(Id\times s(b)\big) K_1(a)=K_1(a)\big(Id\times s(b)\big)$ and using that $s(a)^2 =Id$, one proves that this equation 
is equivalent to the set-theoretical reflection equation for ${\cal R}(a,b)$.
\finprf
Our proposition therefore implies that the reflection maps for the $F_I$ to $F_V$ families are the same as the reflection maps for the 
corresponding families $H_I-H_V$\footnote{With the understanding that $F_{III}$, 
$H_{III}^A$ and $H_{III}^B$ share the same solutions and recalling that $H_{IV}$ does not exist.}. We can therefore restrict our attention
to the $F$ families, which we 
reproduce in Table \ref{families} for completeness. 
\begin{table}[htb]
\label{F_families}
\begin{equation}
\begin{array}{|c|c|c|c|}
\hline
\text{Type} & f_{ab}(X,Y) & g_{ab}(X,Y) & P\\
\hline
&&&\\
F_I & aYP &bXP &\frac{(1-b)X+b-a+(a-1)Y}{b(1-a)X+(a-b)XY+a(b-1)Y}\\
&&&\\
\hline
&&&\\
F_{II} & \frac{Y}{a}P & \frac{X}{b}P &\frac{aX-bY+b-a}{X-Y}\\
&&&\\
\hline
&&&\\
F_{III} & \frac{Y}{a}P & \frac{X}{b}P &\frac{aX-bY}{X-Y}\\
&&&\\
\hline
&&&\\
F_{IV} & YP & XP &1+\frac{b-a}{X-Y}\\
&&&\\
\hline
&&&\\
F_{V} & Y+P & X+P &\frac{a-b}{X-Y}\\
&&&\\
\hline
\end{array}
\nonumber
\end{equation}
\caption{\label{families}Families F for the quadrirational map}
\end{table} Finally, since these families involve parameters, we will work directly with parametric maps.

\subsection{Method}

We propose an adaptation of the folding method that was used in \cite{VZ} to obtain families of reflection maps from the 
YB map associated to the vector nonlinear Schr\"odinger equation. Roughly speaking, the idea is to see a reflection map as a YB map 
modulo some folding applied on the variables on which the YB map acts. More precisely, here we start from Fig. \ref{fig:Rmap}
and impose that $b=\sigma(a)$ for some map $\sigma$ and that $Y=\varphi_a(X)$ and  $V=\varphi_a(U)$ for some map $\varphi_a$ depending on $a$. 
For convenience, we assume that $\sigma$ is an involution as well as $\varphi_a$, the latter meaning that 
\bea
\label{involution_phi}
\varphi_{\sigma(a)}(\varphi_a(X))=X\,.
\eea
Now, since $(U,V)$ is related to $(X,Y)$ by the YB map $\yR(a,b)$, 
we obtain two
possibilities for expressing $V$ in terms of $X$
\bea
\label{U1}
V&=&\varphi_{a}\left(f_{a\sigma(a)}(X,\varphi_a(X))\right)~=~g_{a\sigma(a)}(X,\varphi_a(X))\,.
\eea
This provides a functional equation constraining admissible functions $\varphi_a$ and $\sigma$.
So, for each such admissible pair $(\varphi_a,\sigma$), there is a well-defined map between $V$ and $X$ which we denote $h_a$ 
\bea
\label{U3}
V &=& h_{a}(X)\,,
\eea
and which can be computed using (\ref{U1}). The map $h_a$ together with $\sigma$ are our candidates to get a reflection map $\rR$ as in (\ref{def_ref_map}).

To complete our set of functional equations,
we can also look at the folding method from the point of view 
of the inverse map $\yR^{-1}(a,b)$ which maps $(U,V)$ to $(X,Y)=(g_{ba}(V,U),f_{ba}(V,U))$ 
(due to the reversibility of $\yR$)
and the companion map $\overline{\yR}(b,a)$ 
which maps $(V,X)$ to $(Y,U)\equiv(\bar{f}_{ba}(V,X),\bar{g}_{ba}(V,X))$\footnote{Note that conventions for these maps may vary in the literature but 
results are unaffected by such variations since all four maps $\yR$, $\yR^{-1}$, $\overline{\yR}$, $\overline{\yR}^{-1}$ coincide as shown in 
\cite{ABS2}.}.
From the folding with the inverse map, we get the constraints 
\bea
\label{eq:twu}
Y&=&\varphi_a\big(g_{\sigma(a)a}(\varphi_a(U),U) \big) ~=~ f_{\sigma(a)a}(\varphi_a(U),U)\,.
\eea
Notice that, at first glance, this provides a new constraint for $\varphi$ and $\sigma$.
However, for the quadrirational maps we consider in this paper, listed in Table \ref{families}, 
we have $\yR_{12}(a,b)=\yR_{21}(b,a)$ i.e. $f_{ab}(X,Y)=g_{ba}(Y,X)$. Therefore, the constraint 
given by (\ref{eq:twu}) is equivalent to (\ref{U1}).  
Now, provided that admissible functions  $(\varphi_a, \sigma)$ and the related $h_a$ are found,
we deduce that  $Y=h_a(U)$.
Using this last result and the folding for the companion map $\overline{\yR}(b,a)$, we get 
\bea
\label{Ubar1}
Y&=& h_{a}\left(\bar{g}_{\sigma(a) a}(h_{a} (X),X)\right)~ =~ \bar{f}_{\sigma(a)a} (h_{a} (X),X)\,.
\eea
The quadrirational maps considered in this paper satisfy the following additional property 
\bea
\bar{f}_{ab}(X,Y)=g_{ba}(Y,X)\,,\quad \bar{g}_{ab}(X,Y)=f_{ba}(Y,X)\,.
\eea
Taking account of these properties, one gets
\bea
\label{Ubar2}
Y&=& h_{a}\left(f_{a\sigma(a)}(X,h_{a} (X))\right) ~ =~   g_{a\sigma(a)} (X,h_{a} (X)) \,.
\eea
Comparing (\ref{U1}) and (\ref{Ubar2}), we observe that $\varphi_a$ and $h_a$ play completely dual roles. 
Therefore, if $\varphi_a$ is chosen to be of a certain form, then $h_a$ should be of the same form and vice versa. This is reminiscent of the properties 
derived in \cite{ABS2} concerning the quadrirationality of a YB map, its companion map and their inverses. Also, the involution property 
(\ref{involution_phi}) should hold for $h_a$. 

Combining the ideas of the folding of YB map and the duality between $\varphi_a$ and $h_a$, we are able to  classify  solutions of the set-theoretical reflection equation for the $5$ 
families of quadrirational YB maps presented in \cite{ABS2}. In our context, it is natural to choose $\varphi_a$ and $h_a$ to be both M\"obius maps with 
coefficients being first order polynomials 
in $a$:
\bea
\varphi_a(X)=\frac{p_1(a)X+p_2(a)}{p_3(a)X+p_4(a)}~~,~~h_a(X)=\frac{q_1(a)X+q_2(a)}{q_3(a)X+q_4(a)}\,,
\eea
where $p_j(a)=p_j^0+p_j^1 \,a$ and $p_j(a)=q_j^0+q_j^1 \,a$.
 This corresponds to $16$ free parameters a priori. We also choose $\sigma$ to be an involutive M\"obius map, which gives $3$ additional parameters:
\bea
\sigma(a)=\frac{c_1\,a+c_2}{c_3\,a-c_1}\,.
\eea 
We then solve for these $19$ parameters by inserting into the following constraints as explained above
\bea
\label{constraint1}
h_a(X)=g_{a\sigma(a)}(X, \varphi_{a}(X))\,,\\
\label{constraint2}
\varphi_{\sigma(a)}(\varphi_a(X))=X\,,\quad h_{\sigma(a)}(h_a(X))=X\,.
\eea
The resulting solutions are inserted in the set-theoretical reflection equation to check their validity.

\subsection{Classification}

It is straightforward to check that the identity map (\ie both $h_a$ and $\sigma$ are the identity map) 
is a solution for all families so in the following, we only present nontrivial 
solutions, by which we mean that either $\sigma$ or $h_a$ or both are not the identity map.
With the exception of one solution for the $F_I$ solution, all the maps satisfying (\ref{constraint1}) 
and (\ref{constraint2}) are reflection maps as we found by inspection. 
Note that these operations have been performed with the help of symbolic computation softwares. 
The final results are reported in Table \ref{classification}. 
Note that we did not find any solution for the $F_V$ family. The duality property is explicitly shown in the table: the roles of $h_a$ and $\varphi_a$ 
can be swapped. All solutions depend on an arbitrary parameter $\mu$.
In view of the results in \cite{VZ}, this could be interpreted as the parameter controlling the boundary and, accordingly, the integrable 
boundary conditions of the underlying integrable systems associated to the quadrirational YB maps of the $F$ families.
\begin{table}[htb]
\bea
\begin{array}{|c|c|c|c|}
\hline
\text{Type} & \sigma(a) & \varphi_a(X)~ \text{(resp. $h_a(X)$ )} & h_a(X)~ \text{ (resp. $\varphi_a(X)$ )} \\
\hline
&&&\\
   &\frac{\mu^2}{a} &  \frac{\mu X}{a}  & \frac{(X(1+\mu)-\mu-a)\mu}{X(a+\mu)-(1+\mu)a}\\
   &&&\\
   \hhline{~|-|-|-}
&&&\\
 F_I   &\frac{a+\mu^2-1}{a-1} &  \frac{(a+\mu^2-1)X}{X(a-\mu-1)+a\mu}  & \frac{a+\mu-X\mu-1}{a-1}\\
  &&&\\
   \hhline{~|-|-|-}
&&&\\
   &\frac{1-a}{a(\mu^2-1)+1} &  \frac{(a-1)X}{X(a+\mu a-1)-a\mu}  & \frac{\mu a+1-X\mu-a}{1+a(\mu^2-1)}\\
   &&&\\
\hline
&&&\\
 &\frac{\mu^2}{a}  & \frac{a}{\mu}(X-1)+1 & -\frac{aX}{\mu}  \\
F_{II}& &&\\
\hhline{~|-|-|-}
&&&\\
 &-a+2\mu  & \frac{X}{2X-1} & \frac{a-\mu-aX}{a-2\mu}  \\
& &&\\
\hline
&&&\\
F_{III} & \frac{\mu^2}{a} & \frac{aX}{\mu}  & -\frac{aX}{\mu}\\
&&&\\
\hline
&&&\\
F_{IV} &-a+2\mu & -X & X-a+\mu  \\
&&&\\
\hline
\end{array}\nonumber
\eea
\caption{\label{classification}Reflection maps for quadrirational map $F_I-F_{IV}$}
\end{table}

In the geometric construction of the quadrirational maps $F_I-F_V$ in \cite{ABS2}, the so-called singular points play an important role.
It turns out that we can make use of these points in our construction to obtain other reflection maps which we call degenerate. This 
is achieved by allowing $\varphi_a(X)$ in (\ref{U1}) to take on the singular 
values of the corresponding family. In general, this 
can be $0,1,\infty,\sigma(a)$. Each time a degenerate case
is solution, the reflection equation is satisfied for \textit{any} map $\sigma$. 
The duality property still holds for these degenerate 
cases. We present the results in Table \ref{degenerate_maps}.
\begin{table}[ht]
\bea
\begin{array}{|c|c|c|}
\hline
\text{Type}  & h_a(X)~ \text{ (resp. $\varphi_a(X)$ )} & \varphi_a(X)~ \text{(resp. $h_a(X)$ )}\\
\hline
&&\\
    & \frac{\sigma(a)(a-1)X}{a(X-1)+\sigma(a)(a-X)} &\infty\\
    &&\\
F_I  & \frac{a-X+\sigma(a)(X-1)}{a-1}&0\\
&&\\
    & \frac{\sigma(a)}{a}X &1\\
    &&\\
    &X &\sigma(a)\\
&&\\
\hline
&&\\
    & X &\infty\\
    &&\\
F_{II}  & \frac{a}{\sigma(a)}(X-1)+1 &0\\
&&\\
  & \frac{a}{\sigma(a)}X &1\\
&&\\
\hline
&&\\
  & X &\infty\\
F_{III}&&\\
  &\frac{a}{\sigma(a)}X &0\\
&&\\
\hline
&&\\
  & X  &\infty\\
F_{IV}&&\\
&X-a+\sigma(a)&0\\
&&\\
\hline
&&\\
F_{V}  & X & \infty\\
&&\\
\hline
\end{array}
\nonumber
\eea
\caption{\label{degenerate_maps}Degenerate reflection maps}
\end{table}

\section*{Conclusions and perspectives}

We introduced the general theory of the set-theoretical reflection equation. 
The associated notion of reflection maps
originally introduced in \cite{VZ} has been connected to the concept of quadrirational maps 
which is itself strongly related to the so-called 
Yang-Baxter maps and the idea of 3D consistency of equations on quad-graphs. We have used the quadrirational YB maps obtained in \cite{ABS2,PSTV} to 
derive a classification of reflection maps associated to them, thus providing a large family of such maps in addition to the two classes 
originally found by two of the authors in \cite{VZ} from the study of factorization of vector soliton interactions in the presence of an 
integrable boundary. Some ideas of the latter paper were adapted here to perform a rather "brute force" construction. It would be nice to 
reinterpret our results in more geometrical terms along the lines of the use of results on pencils of conics which led to the classification 
in \cite{ABS2}. More generally, one can also wonder how the general and abstract construction of YB maps as performed in \cite{WX,ESS} can be adapted to 
yield a general construction of the associated reflection maps.

The results of this paper open the way to a wider programme of understanding reflection maps in the context of quad-graph systems and 
fully discrete integrable systems. Indeed, just like YB maps are one aspect of the idea of discrete integrability related to the 3D 
consistency approach, one can ask the important question of the meaning of a reflection map in the context of 3D consistency on a 
quad-graph. We are currently investigating this challenging issue and will report our results in a forthcoming paper. When this is understood, 
it will be interesting to see if the interpretation of a reflection map as an object encoding the integrable interaction between a soliton and a 
boundary which was found in \cite{VZ} in the context of the vector nonlinear Schr\"odinger equation also holds in the realm of fully discrete integrable systems.
We also believe that reflection maps can have a wide range of applications, for example in quantum algebra, in geometric crystals 
or in the soliton cellular automaton.

It seems also quite plain to us that different generalizations of the set-theoretical reflection equation introduced in this paper are possible,
mimicking existing generalizations of the quantum reflection equation. Indeed, after the introduction of the quantum reflection equation in 
\cite{Che,Sk}, a more general structure, known as quadratic algebras, was introduced in \cite{FM}: the structure that is considered in that paper 
involves four different matrices subject to general Yang-Baxter like equations. It could be interesting to study the 
set-theoretical analog of these quadratic algebras. Following the same idea, it may be possible to define
a reflection type equation associated to the dynamical set-theoretical Yang-Baxter equation \cite{Shi}. In the quantum context, this type of equation has been studied 
e.g. in \cite{FHS97}.


\begin{thebibliography}{99}
\bibitem{yang} C.N.Yang, 
\textsl{Some exact results for the many-body problem in 
one dimension with repulsive delta-function interaction},
  Rev.Lett. \textbf{19} (1967) 1312.

\bibitem{baxter} R.J.Baxter,
  \textsl{Partition function of the eight-vertex lattice model,}
  Ann.Phys. \textbf{70} (1972) 193; 
  \textsl{Asymptotically degenerate maximum eigenvalues of the eight-vertex 
  model transfer matrix and interfacial tension},
J.Stat.Phys. \textbf{8} (1973) 25; 
  \textsl{Exactly solved models in statistical mechanics} (Academic 
Press, 1982).

\bibitem{Drin} V.G.Drinfeld, 
\textsl{On some unsolved problems in quantum group theory,} Quantum groups
(Leningrad, 1990), Lecture Notes in Mathematics, edited by P. P. Kulish (Springer Verlag,
Berlin, 1992), Vol. 1510, pp. 1-8.

\bibitem{Skly88} E.K.Sklyanin,
\textsl{Classical limits of $SU(2)$-invariant solutions of the Yang-Baxter solution.}
J.Soviet.Math. \textbf{40} (1988) 93.

\bibitem{BK} A.Berenstein and D.Kazhdan,
\textsl{Geometric and unipotent crystals,}
GAFA special volume (2000) 188 and \texttt{arXiv:math/9912105}.

\bibitem{Eti} P.Etingof,
\textsl{Geometric crystals and set-theoretical solutions to the quantum Yang-Baxter equations,}
Commun.algebra \textbf{31} (2003) 1961 and \texttt{arXiv:math/0112278}.

\bibitem{TS} D.Takahashi and J.Satsuma,
\textsl{A soliton cellular automaton,}
J.Phys.Soc.Japan \textbf{59} (1990) 3514.

\bibitem{HKT} G.Hatayama, A.Kuniba and T.Takagi,
\textsl{Soliton cellular automata associated with crystal bases,}
Nucl.Phys. \textbf{B577} (2000) 619 and \texttt{arXiv:solv-int/9907020}. 

\bibitem{Gon} V.M.Goncharenko, 
\textsl{Multisoliton solutions of the matrix KdV equation},
Theor. Math. Phys. {\bf 126} (2001) 81.

\bibitem{Tsu} T.Tsuchida, 
\textsl{N-Soliton Collision in the Manakov Model,}
Prog.Theor.Phys. \textbf{111} (2004) 151 and \texttt{arXiv:nlin/0302059}.

\bibitem{APT} M.J.Ablowitz, B.Prinari and A.D.Trubatch, 
\textsl{Soliton interactions in the vector NLS equation,} 
Inverse Problems \textbf{20} (2004) 1217.

\bibitem{ABS}V.E.Adler, A.I.Bobenko and Yu.B.Suris,
\textsl{Classification of integrable equations on quad-graphs. The consistency approach,}
Commun.Math.Phys. \textbf{233} (2003) 513 and \texttt{arXiv:nlin/0202024}.

\bibitem{PTV}V.G.Papageorgiou, A.G.Tongas and A.P.Veselov,
\textsl{Yang-Baxter maps and symmetries of integrable equations on quad-graphs,} 
J.Math.Phys. \textbf{47} (2006) 083502 and \texttt{arXiv:math/0605206}.

\bibitem{Ves} A.P.Veselov, 
\textsl{Yang-Baxter maps and integrable dynamics,}
Phys.Lett. {\bf A314} (2003) 214 and \texttt{arXiv:math/0205335}.

\bibitem{SV} Y.B.Suris and A.P.Vesolv,
\textsl{Lax matrices for Yang-Baxter maps,}
J.Nonlin.Math.Phys. \textbf{10} (2003) 223 and \texttt{arXiv:math/0304122}. 

\bibitem{ABS2}V.E.Adler, A.I.Bobenko and Yu.B.Suris,
\textsl{Geometry of Yang--Baxter maps: pencils of conics and quadrirational mappings,}
Commun.Anal.Geom. {\bf 12} (2004) 967 and \texttt{arXiv:math/0307009}.

\bibitem{PSTV}V.G.Papageorgiou, Yu.B.Suris, A.G.Tongas and A.P.Veselov,
\textsl{On quadrirational Yang-Baxter Maps,} 
SIGMA {\bf 6} (2010) 033 and \texttt{arXiv:0911.2895}.
 
\bibitem{Che} I.V.Cherednik,
\textsl{Factorizing particles on a half line and root systems,}
Theor.Math.Phys. \textbf{61} (1984) 977 and Teor.Mat.Fiz. \textbf{61} (1984) 35.

\bibitem{Sk}E.K.Sklyanin,
\textsl{Boundary conditions for integrable quantum systems,}
J.Phys. \textbf{A21} (1988) 2375.

\bibitem{KS} P.P.Kulish and E.K.Sklyanin,
\textsl{Algebraic structures related to the reflection equations,}
J.Phys. \textbf{A25} (1992) 5963 and \texttt{hep-th/9209054}.

\bibitem{MNO}A.Molev, M.Nazarov and G.Olshansky,
\textsl{Yangians and classical Lie algebras,}
Russ.Math.Surveys \textbf{51} (1996) 205 and \texttt{hep-th/9409025}.

\bibitem{BPS}D.Bernard, V.Pasquier and D.Serban,
\textsl{Exact solution of long range interacting spin chains with boundaries,}
 \texttt{hep-th/9501044}.

\bibitem{CC}V.Caudrelier and N.Crampe,
\textsl{Integrable N-particle Hamiltonians with Yangian or reflection algebra
symmetry,} 
J.Phys. \textbf{A37} (2004) 6285 and \texttt{math-ph/0310028}. 

\bibitem{KOY}A.Kuniba, M.Okado and Y.Yamada,
\textsl{Box-ball system with reflecting end,}
J.Nonlin.Math.Phys. \textbf{12} (2005) 475 and \texttt{nlin/0411044}.

\bibitem{HL}M.Halln\"as and E.Langmann,
\textsl{Exact solutions of two complementary 1D quantum many-body systems on the half-line,}
J.Math.Phys. \textbf{46} (2005) 052101 and \texttt{math-ph/0404023}.

\bibitem{bart}V.Caudrelier and N.Crampe,
\textsl{Exact results for the one-dimensional many-body problem with contact
interaction: Including a tunable impurity,}
Review Math.Phys. \textbf{19} (2007) 349 and \texttt{cond-mat/0501110}. 

\bibitem{VZ} V.Caudrelier and Q.C.Zhang, 
\textsl{Yang-Baxter and reflection maps from vector solitons with a boundary,} 
\texttt{arXiv:1205.1133}.

\bibitem{VZ2} V.Caudrelier and Q.C.Zhang, \textsl{Vector Nonlinear Schr\"odinger Equation on the half-line,} 
J.Phys. {\bf A45} (2012) 105201 and \texttt{arXiv:1110.2990}.

\bibitem{Skly_classical} E.K.Sklyanin, \textsl{Boundary conditions for integrable equations,} Funct.Anal.App. {\bf 21}
(1987) 164. 
 
\bibitem{H} I.T.Habibullin, \textsl{Integrable initial-boundary value problems,} Theor.Math.Phys. {\bf 86} (1991) 28;
P. N. Bibikov, V. O. Tarasov, \textsl{A boundary-value problem for the nonlinear Schr\"odinger
equation,} Theor.Math.Phys. {\bf 79} (1989) 334.

\bibitem{GGH} B.G\"urel, M.G\"urses and I.T.Habibullin, \textsl{Boundary value problems for integrable
equations compatible with the symmetry algebra,} J.Math.Phys. {\bf 36} (1995) 6809 and \texttt{solv-int/9411001}.

\bibitem{Fokas} A.S.Fokas, \textsl{A Unified Approach to Boundary Value Problems}, CBMS-SIAM (2008).

\bibitem{AK}C.Ahn and W.M.Koo,
\textsl{Boundary Yang-Baxter equation in the RSOS representation,}
Statistical models, Yang-Baxter equation
and related topics, Tianjin (1995)  3-12 and
\texttt{hep-th/9508080}. 

\bibitem{FHS} 
H.Fan, B.-Y.Hou and K.-J.Shi,
\textsl{General solution of reflection equation for eight-vertex SOS model,}
J.Phys. \textbf{A28} (1995) 4743.

\bibitem{BPO} R.E.Behrend, P.A.Pearce and D.L.O'Brien,
\textsl{Interaction-Round-a-Face models with fixed boundary conditions: The ABF fusion hierarchy,}
 J.Stat.Phys. \textbf{84} (1996) 1 and \texttt{arXiv:hep-th/9507118}.

\bibitem{Bax86} R.J.Baxter,
\textsl{The Yang-Baxter equations and the Zamolodchikov Model,} 
Physica \textbf{18D} (1986) 321.

\bibitem{BMS}V.V.Bazhanov, V.V.Mangazeev and S.M.Sergeev,
\textsl{Quantum geometry of 3-dimensional lattices,}
 J.Stat.Mech.0807:P07004,2008 and \texttt{arXiv:0801.0129}.

\bibitem{WX} A.Weinstein and P.Xu,
\textsl{Classical solutions of the Yang-Baxter equation,}
Commun.Math.Phys. \textbf{48} (1992) 309.

\bibitem{ESS} P.Etingof, T.Schedler and A.Soloviev,
\textsl{Set-theoretical solutions of the quantum Yang-Baxter equation,}
Duke Math.J. \textbf{100} (1999) 169 and \texttt{arXiv:math/9801047}.

\bibitem{FM}L.Freidel and J.M.Maillet,
\textsl{Quadratic algebras and integrable systems,}
Phys.Lett. \textbf{B262} (1991) 278.

\bibitem{Shi} Y.Shibukawa, 
\textsl{Dynamical Yang-Baxter maps,} 
Int.Math.Res.Notices \textbf{36} (2005);
\textsl{Dynamical Yang-Baxter maps with an invariance condition,} 
Publ.Res.Inst.Math.Sci. \textbf{43} (2007) 1157 and \texttt{arXiv:0704.3109}.
    
\bibitem{FHS97}
H.Fan, B.-Y.Hou and K.Shi,
\textsl{Representation of the boundary elliptic quantum group
$BE_{\tau,\eta}(sl_2)$ and the Bethe ansatz}, 
Nucl.Phys. \textbf{B496} (1997) 551.
    
\end{thebibliography}
\end{document}